\documentclass{svjour3}                     
\smartqed  
\usepackage{graphicx}
\usepackage{url}
\usepackage{amsfonts}
\journalname{Journal of Automated Reasoning}

\begin{document}

\title{Computing with Classical Real Numbers}

\author{Cezary Kaliszyk \and Russell O'Connor}

\institute{Radboud University Nijmegen \\
              \email{\{cek,roconnor\}@cs.ru.nl} 
}

\date{Received: date / Accepted: date}

\maketitle

\begin{abstract}
There are two incompatible Coq libraries that have a theory of the real
numbers; the Coq standard library gives an axiomatic treatment of
classical real numbers, while the CoRN library from Nijmegen defines
constructively valid real numbers. 
Unfortunately, this means results about one structure cannot easily be used in the other structure.
We present a way interfacing these two libraries by showing that their real number structures are isomorphic assuming the classical axioms already present in the standard library reals.
This allows us to use O'Connor's decision procedure for solving ground inequalities
present in CoRN to solve inequalities about the reals from the Coq standard library, and it allows theorems from the Coq standard library to apply to problem about the CoRN reals.
\end{abstract}

\section{Introduction}\label{intro}

Coq is a software proof assistant based on dependent type theory developed at INRIA~\cite{Coq:2007}.
By default, it reasons with constructive logic via the Curry-Howard isomorphism.
The Curry-Howard isomorphism associates propositions with types and proofs of propositions with programs of the associated type.
This makes Coq a functional programming language as well as a deduction system.
The identification of a programming language with a deduction system allows Coq to reason about programs and allows Coq to use computation to prove theorems.

Coq can support classical reasoning by the declaration of additional axioms; however, these additional axioms will not have any corresponding computational component.
Therefore, using axioms may limit the ability for Coq to use computation to prove theorems.

At least two different developments of the real numbers have been created for Coq.
Coq's standard library declares the existence of the real numbers in an axiomatic fashion.
This library also requires the axioms for classical logic. It gives users the familiar, classical real numbers.

The other formalization of the real numbers is done constructively in the CoRN library~\cite{lcf04-corn}.
This library specifies what a constructive real number structure is, and proves that all such real number structures are isomorphic.
These real numbers are constructive and there is one effective implementation where real numbers can be evaluated to arbitrary precision within Coq.

In this paper we show how to connect these two developments by showing that Coq's real numbers form a real number structure in CoRN.  We do this by

\begin{itemize}
\item Deriving some logical consequences of the classical real numbers (Section \ref{p5:sec:logic}).  Specifically, we formally prove the well-known result that $\Pi^1_0$ would be decidable.
\item Using these logical consequences to prove that the classical real numbers form a constructive real number structure (Section \ref{p5:sec:iso}).
\item Using the resulting isomorphism between classical and constructive real numbers to prove some classical real number inequalities by evaluating constructive real number expressions (Section \ref{p5:sec:computation}).
\end{itemize}

\subsection{The two universes of Coq}\label{intro-prop-set}

Coq has a mechanism for program extraction~\cite{Letouzey02}.
Programs developed in Coq can be translated into Ocaml, Haskell, or Scheme.
If these programs are proved correct in Coq, then the extracted programs have high assurance of correctness.

To facilitate extraction, Coq has two separate universes: the \verb|Set| universe, and the \verb|Prop| universe (plus an infinite series of \verb|Type| universes on top of these two).
The \verb|Prop| universe is intended to contain only logical propositions and its values are discarded during extraction.
The types in the \verb|Set| universe are computationally relevant the values of these types make up the extracted code.
In order to maintain the soundness of extraction, the type system prevents information from flowing from the \verb|Prop| universe to the \verb|Set| universe.
Otherwise, vital information could be thrown away during extraction, and the extracted programs would not run.

The \verb|Prop|/\verb|Set| distinction will play an important role in our work.
The logical operators occur in both universes.  The following table lists some logical operations and their corresponding syntax for both the \verb|Prop| and \verb|Set| universes.

\begin{table}[h]
\begin{tabular}{|l|l|l|}
  \hline
  Math Notation & \verb|Prop| Universe & \verb|Set| Universe\\
  \hline
  $A \wedge B$ & \verb|A /\ B| & \verb|A * B|\\
  \hline
  $A \vee B$ & \verb|A \/ B| & \verb|A + B|\\
  \hline
  $A \rightarrow B$ & \verb|A -> B| & \verb|A -> B|\\
  \hline
  $\neg A$ & \verb|~A| & \textit{not used}\\
  \hline
  $\forall x : X. P (x)$ & \verb|forall x:X, P x| & \verb|forall x:X, P x|\\
  \hline
  $\exists x : X. P (x)$ & \verb|exists x:X, P x| & \verb!{ x : X | P x }!\\
  \hline
\end{tabular}
\end{table}

One might think that proving that classical real numbers satisfy the requirements of a constructive real number structure would be trivial.  Should not the constructive 
requirements be no stronger than the classical requirement for a real number structure when we use classical reasoning?  However,
Coq's \verb|Prop|/\verb|Set| distinction prevents a naive attempt at creating such an isomorphism between the classical and constructive real numbers.
The difficulty is that classical reasoning is only allowed in the \verb|Prop| universe.
A constructive real number structure requires a \verb|Set|-style existence in the proof that the a sequence converges to its limit (see Section \ref{p5:sec:iso-building}), but the theory provided by the Coq standard library only proves a classical \verb|Prop|-style existence.

Although we have recently discovered a direct way to show that the classical Coq reals form a constructive real number structure, we will present our original solution that transforms the classical existentials provided by the Coq standard library into a constructive existential.
This solution uses Coq's real number axioms to create constructive existentials from classical existentials for any $\Pi^1_0$ sentence (Section \ref{p5:sec:logic}), a result that has wide reaching applications.


\section{Logical Consequences of Coq real numbers}\label{p5:sec:logic}

Coq's standard library defines the classical real numbers axiomatically.
This axiomatic definition has some general logical consequences.
In this section we present some of the axioms used to define the real numbers and then show how they imply the decidability of $\Pi_0^1$ sentences.
The axioms of the real numbers cannot be effectively realized, so a decision procedure for $\Pi_0^1$ sentences is not implied by this decidability result.

\subsection{The axiomatic definition of the real numbers}

The definition for the reals in the Coq standard library asserts a set
$\mathbb{R}$, the constants $0$, $1$, and the basic arithmetic operations:

\begin{verbatim}
Parameter R : Set.
Parameter R0 : R.
Parameter R1 : R.
Parameter Rplus : R -> R -> R.
Parameter Rmult : R -> R -> R.
...
\end{verbatim}
A numeric literal, such as \verb|20|, is simply notation for an expression, such as:

\begin{verbatim}
(R1+R1)*((R1+R1)*(R1+(R1+R1)*(R1+R1)))
\end{verbatim}

In addition to the arithmetic operations, an order relation is asserted.
\begin{verbatim}
Parameter Rlt : R -> R -> Prop.
\end{verbatim}

Axioms for these operations and relations define their semantics.
There are 17 axioms.
We show only some relevant ones; the entire list of axioms can be found in the Coq standard library.
The properties described by the axioms include associativity
and commutativity of addition and multiplication, distributivity, and neutrality of zero and
one.

\begin{verbatim}
Axiom Rplus_comm : forall r1 r2:R, r1 + r2 = r2 + r1.
...
\end{verbatim}

There are also several axioms that state that the order relation for the real numbers form a total order.
The most important axiom for our purposes will be the law of trichotomy:

\begin{verbatim}
Axiom total_order_T : forall r1 r2:R, {r1 < r2} + {r1 = r2} + {r1 > r2}.
\end{verbatim}

Finally, there is an Archimedian axiom and an axiom stating the least upper bound property.

\begin{verbatim}
Parameter up : R -> Z.
Axiom archimed : forall r:R, IZR (up r) > r /\ IZR (up r) - r <= 1.

Axiom completeness :
    forall E:R -> Prop,
      bound E -> (exists x : R, E x) -> sigT (fun m:R => is_lub E m).
\end{verbatim}

\subsection{Decidability of $\Pi_0^1$ sentences}

It is important to notice that the trichotomy axiom uses \verb|Set|-style disjunctions.
This means that users are allowed to write functions that make decisions by comparing real numbers.
This axiom might look surprising at first since real numbers are infinite structures and therefore comparing them is impossible in finite time in general.
The motivation for this definition comes from classical mathematics where mathematicians regularly create functions based on real number trichotomy.
It allows one to define a step function, which is not be definable in constructive mathematics.

This trichotomy axiom can be used to decide any $\Pi_0^1$ property.
For any decidable predicate over natural numbers $P$, we first define a sequence of terms that take values when the property is true:

\begin{equation}\displaystyle
a_n =_{\mathrm{def}} \left\{ \begin{array}{ll}
\frac{1}{2^n} & \textrm{if } P(n)\textrm{ holds}\\
0             & \textrm{otherwise} \\
\end{array} \right.
\end{equation}

We can then define the sum of this infinite sequence, which is guaranteed to converge:

\begin{equation}
\displaystyle S =_{\mathrm{def}}
  \sum_{n=0}^{\infty}a_n
\end{equation}

The trichotomy axiom allows us to compare $S$ with $2$.
It follows that if $\displaystyle S = 2$ then the predicate $P$ hold for every natural number, and if $\displaystyle S < 2$ then it is not the case (the case of $S > 2$ is easily ruled out).
Furthermore, this distinction can be made in \verb|Set| universe.

We formalized the above reasoning in Coq and we obtained the following logical theorem.
\begin{verbatim}
  forall_dec
     : forall P : nat -> Prop,
       (forall n : nat, {P n} + {~ P n}) ->
       {(forall n : nat, P n)} + {~ (forall n : nat, P n)}
\end{verbatim}

\subsubsection{Constructive indefinite description}

We can extend the previous result by using a general logical lemma of Coq.
The constructive indefinite description lemma states that if we have a decidable predicate over the natural numbers, then we can convert a \verb|Prop| based existential into a \verb|Set| based one.
Its formal statement can be found in the standard library:

\begin{verbatim}
  constructive_indefinite_description_nat
     : forall P : nat -> Prop,
       (forall x : nat, {P x} + {~ P x}) ->
       (exists n : nat, P n) -> {n : nat | P n}
\end{verbatim}

This lemma can be seen as a form of Markov's principle in Coq.
The lemma works by doing a bounded search for a new witness satisfying the predicate.  The witness from the \verb|Prop| based existential is only used to prove
termination of the search.
No information flows from the \verb|Prop| universe to the \verb|Set| universe because the witness found for the \verb|Set| based existential is independent of the witness from the \verb|Prop| based one.

Classical logic allows us to convert a negated universal statement into an existential statement in \verb|Prop|:

\begin{verbatim}
  not_all_ex_not
     : forall (U : Type) (P : U -> Prop),
       ~ (forall n : U, P n) -> exists n : U, ~ P n
\end{verbatim}

By combining these theorems with our previous result, we get a theorem whose conclusion is either a constructive existential or a universal statement:

\begin{verbatim}
  sig_forall_dec
     : forall P : nat -> Prop,
       (forall n : nat, {P n} + {~ P n}) ->
       {n : nat | ~ P n} + {(forall n : nat, P n)}
\end{verbatim}

To see why this implies all $\Pi_0^1$ sentences are decidable, consider an arbitrary $\Pi_0^1$ sentence $\varphi$.  If $\varphi$ is $\Pi_0^0$, then it is decidable by the basic
properties of $\Pi_0^0$ sentences. Otherwise, if $\varphi$ is of the form $\forall n:\mathbb{N}. \psi(n)$ where $\psi(n)$ is decidable by our inductive hypothesis. The above lemma allows use to conclude that $\varphi$ is decidable from the fact that $\psi(n)$ is decidable.

\section{The construction of the isomorphism}\label{p5:sec:iso}

In this section we briefly present the algebraic hierarchy present in CoRN
(it is described in detail in \cite{cornhierarchy} and \cite{luisthesis}).
We show that the Coq reals fulfill the requirements of a constructive real number structure, and hence they are isomorphic to any other real number structure.

\subsection{Building a constructive reals structure based on Coq reals}\label{p5:sec:iso-building}

The collection of properties making up a real number structure in CoRN is broken down to form a hierarchy of different structures.
The first level, \verb|CSetoid|, defines the properties for equivalence and apartness.  The next level is \verb|CSemigroup| which defines some properties for addition.
More structures are defined on top of each other all the way up to the \verb|COrderedField| structure---a constructive ordered field does not require trichotomy.
Finally, the \verb|CReals| structure is defined on top of the \verb|COrderedField| structure.
The full list of structures is given below.

\begin{tabular}{l@{ \qquad--\qquad }l}
 CSetoid & constructive setoid \\
 CSemiGroup  & semi group \\
 CMonoid & monoid \\
 CGroup & group \\
 CAbGroup & Abelian group \\
 CRing & ring \\
 CField & field \\
 COrdField & ordered field \\
 CReals & real number structure\\
\end{tabular}\\

To prove that classical reals form a constructive real number structure, we 
created instances of all these structures for the classical real numbers (called \verb|RSetoid|, \verb|RSemigroup|, etc.).
For example, \verb|RSetoid| is the constructive setoid based on Coq real numbers.
The carrier is \verb|R| and Leibniz equality and its negation are used as the equality and apartness relations.
The proofs of the setoid properties of these relations are simple.

The basic arithmetic operations from Coq real numbers shown to satisfy all
the properties required up to \verb|COrdField|.  The proofs of these properties follow
straightforwardly from similar properties provided by Coq's standard library.
For details, we refer the reader to CoRN source files \cite{cornsrc}.
We present just the final step, the creation of the \verb|CReals| structure
based on the ordered field.

Two additional operations are required to form a constructive real numbers structure
from a constructive ordered field: the limit operation and a function that realizes the Archimedian property.
The limit operation is the only step where the facts about Coq reals cannot naively be used to instantiate the required properties.
This is because the convergence property of limits for the Coq reals only establishes that there exists a point where the sequence gets close to the limit using the \verb|Prop| based  quantifier, whereas \verb|CReals| requires such a point to exist using the \verb|Set| based quantifier.
One cannot directly convert a \verb|Prop| based existential into a \verb|Set| based one, because information is not allowed to flow from the \verb|Prop| universe to the \verb|Set| universe.

The exact Coq goal asks us to show that if for any $\epsilon$ there is an index in
a sequence $N$ such that all further elements in the sequence are closer
to the limit value than $\epsilon$.
The related property from the Coq library is shown as hypothesis \verb|u|.

\begin{verbatim}
e : R
e0 : 0 < e
u : forall eps : R,  eps > 0 -> exists N : nat,
    forall n : nat,
    (n >= N)%nat -> Rfunctions.R_dist (s n) x < eps
______________________________________(1/1)
{N : nat | forall m : nat,
  (N <= m)%nat -> AbsSmall e (s m[-]x)}
\end{verbatim}

To prove this goal, we first use the
\verb|constructive_indefinite_description_nat| lemma to reduce the \verb|Set| based existential to a \verb|Prop| based one.
Applying this lemma to our goal from before reduces the problem to the following goal.

\begin{verbatim}
e : R
e0 : 0 < e
u : forall eps : R,  eps > 0 -> exists N : nat,
    forall n : nat,
    (n >= N)%nat -> Rfunctions.R_dist (s n) x < eps
______________________________________(2/2)
exists N : nat, forall m : nat,
  (N <= m)%nat -> AbsSmall e (s m[-]x)}
\end{verbatim}
This now follows easily from the hypothesis.
However, we are also required to prove the decidability of the predicate:


\begin{verbatim}
______________________________________(1/2)
{(forall m : nat, (x0 <= m)%nat -> AbsSmall e (s m[-]x))} +
{~ (forall m : nat, (x0 <= m)%nat -> AbsSmall e (s m[-]x))}
\end{verbatim}

This goal appears hopeless at first because we are required to prove the decidability of a $\Pi_1^0$ sentence.
However, we can use the \verb|forall_dec| lemma from the previous section to prove the decidability of this sentence.
This complete the proof that the classical real numbers form a constructive real number structure.

\subsection{The isomorphism}

Niqui shows in Section 1.4 of his PhD thesis
\cite{miladsthesis} that all constructive reals structures are
isomorphic, the proof is present in CoRN as \texttt{iso\_CReals}.
The constructed isomorphism defines two maps that are inverses of each other and
proves that the isomorphism preserves the constants $0$ and $1$, arithmetic operations
and limits. More details can be found in \cite{miladsthesis}. 

In order to use the isomorphism in an effective way, we need
to show that the definitions of basic constants and the operations are
preserved. Since the reals of the standard library of Coq are written
as \verb|R| and CoRN reals as \verb|IR|, we called the two
functions of the isomorphism \verb|RasIR| and \verb|IRasR|. From
Niqui's construction, one obtains the basic properties of this
isomorphism:


 \begin{itemize}
 \item Preserves equality and inequalities:
\begin{verbatim}
Lemma R_eq_as_IR : forall x y, (RasIR x [=] RasIR y -> x = y).
Lemma IR_eq_as_R : forall x y, (x = y -> RasIR x [=] RasIR y).
Lemma R_ap_as_IR : forall x y, (RasIR x [#] RasIR y -> x <> y).
Lemma IR_ap_as_R : forall x y, (x <> y -> RasIR x [#] RasIR y).
Lemma R_lt_as_IR : forall x y, (RasIR x [<] RasIR y -> x < y).
...
\end{verbatim}

 \item Preserves constants: $0$, $1$ and basic arithmetic operations: $+$, $-$, $*$:
\begin{verbatim}
Lemma R_Zero_as_IR : (RasIR R0 [=] Zero).
Lemma R_plus_as_IR : forall x y, (RasIR (x+y) [=] RasIR x [+] RasIR y).
...
\end{verbatim}
 \end{itemize}

An important difference between the definition of real numbers in the Coq
standard library and in CoRN is the way partiality is handled. Partial
functions are defined as total function for the Coq reals, but their properties
require proofs that the function parameters are in the appropriate
domain. For example, division is defined as a total operation on real
numbers; however, all the axioms that specify properties of division
have assumptions that the reciprocal is not zero.  This
means that the term $\frac{1}{0}$ is some real number, but it is not
possible to prove which one it is.

In CoRN, partial functions require an additional argument, the domain
condition.  Division is a three argument operation; 
the third argument is a proof that the divisor is apart from
zero. Other partial functions, such as the logarithm, are defined in a similar way.
We prove that this isomorphism preserves these partial functions.
These lemmas require a proof that the arguments are in the proper domain to be passed to the domain conditions of the CoRN functions.

\begin{itemize}
 \item Preserves the reciprocal and division for any proof:
\begin{verbatim}
Lemma R_div_as_IR : forall x y (Hy : Dom (f_rcpcl' IR) (RasIR y)), 
  (RasIR (x/y) [=] (RasIR x [/] RasIR y [//] Hy)).
\end{verbatim}
\end{itemize}

Niqui's theorem proves the basic arithmetic operations and limits are preserved by
the isomorphism.
However, the real number structure does not specify any transcendental functions.
Therefore it is necessarily to manually prove that these functions are preserved by the isomorphism.
This may be easy if the Coq and CoRN definitions are similar, but they may be difficult if the two systems choose different definitions for the same function.

\begin{itemize}
 \item Preserves infinite sums:\\
   The proof that the values of the sums are the same requires
   again uses the decidability of $\Pi_0^1$ sentences
   and \texttt{constructive\_indefinite\_description\_nat}.
   The term \verb|prf| is the proof that the series converges.
\begin{verbatim}
Lemma R_infsum_as_IR : forall (y: R) a,
  Rfunctions.infinit_sum a y -> forall prf,
  RasIR y [=] series_sum (fun i : nat => RasIR (a i)) prf.
\end{verbatim}
 \item Preserves transcendental functions: $exp$, $sin$, $cos$, $tan$, $ln$
\begin{verbatim}
Lemma R_exp_as_IR : forall x, RasIR (exp x) [=] Exp (RasIR x).
Lemma R_sin_as_IR : forall x, RasIR (sin x) [=] Sin (RasIR x).
Lemma R_cos_as_IR : forall x, RasIR (cos x) [=] Cos (RasIR x).
Lemma R_tan_as_IR : forall x dom, RasIR (tan x) [=] Tan (RasIR x) dom.
Lemma R_ln_as_IR : forall x dom, RasIR (ln x) [=] Log (RasIR x) dom.
\end{verbatim}
\end{itemize}

We finally prove that the isomorphism preserves the constant $\pi$.
This was more difficult because the $\pi$ in Coq is defined as the infinite sum

\begin{equation}\displaystyle
\pi_{Coq} =_{\mathrm def} \sum_{i=0}^\infty{\frac{(-1)^i}{2 * i + 1}},
\end{equation}
while in CoRN $\pi$ is defined as the limit of the sequence

\begin{equation}\displaystyle
pi_n =_{\mathrm{\mathrm def}} \left\{ \begin{array}{ll}
0             & \textrm{if } n = 0\\
pi_{n-1} + cos(pi_{n-1}) & \textrm{otherwise} \\
\end{array} \right.
\end{equation}

\begin{equation}\displaystyle
\pi_{CoRN} =_{\mathrm{\mathrm def}} \lim_{n \to \infty} pi_n.
\end{equation}

Both libraries contain proofs that the sine of $\pi$ is equal to zero, and
additionally that it is the smallest positive number with this property.
Using these properties it is possible to show that indeed the two definitions
describe the same number:

\begin{verbatim}
  Lemma R_pi_as_IR : RasIR (PI) [=] Pi.
\end{verbatim}


\section{Computation with classical reals}\label{p5:sec:computation}

\subsection{Solving ground inequalities}

O'Connor's work on fast real numbers in CoRN includes a semi-decision
procedure (a decision procedure that may not terminate) for solving
strict inequalities on constructive real numbers.
With the isomorphism it is possible to use it to solve some goals for classical reals.

Consider the example of proving $exp(\pi) - \pi < 20$ for the classical real numbers:

\begin{verbatim}
______________________________________(1/1)
exp PI - PI < 20
\end{verbatim}

Our tactic first converts the Coq inequality to a CoRN inequality by using the fact that the isomorphism preserves inequalities.
Then it recursively applies the facts about the
isomorphism to convert the Coq terms on both sides of the inequality and their corresponding CoRN terms.
This is done with using a rewrite database and
the autorewrite mechanism for setoids.
The advantage of using a rewrite database is that it can be easily extended with
new facts about new functions being preserved under the isomorphism.
The disadvantage of this method is that
the setoid rewrite mechanism is fairly slow in Coq 8.1.

\begin{verbatim}
______________________________________(1/1)
Exp Pi[-]Pi[<](One[+]One)[*]
              ((One[+]One)[*](One[+](One[+]One)[*](One[+]One)))
\end{verbatim}
(Recall that, in Coq, the real number $20$ is simply notation for $(1+1)*((1+1)*(1+(1+1)*(1+1)))$.)

Once the expression is converted to a CoRN expression, CoRN's semi-decision procedure can be applied (which itself uses another rewrite database to change the representation again).
This semi-decision procedure may not terminate.
If the two sides of the inequality are different, it will approximate the real numbers 
accurately enough to either prove the required inequality (or fail if the inequality the other direction holds).
If the two sides are equal, then the search for a sufficient approximation will never terminate.


The decision procedure for CoRN takes an argument which is used for the starting precision of the approximation.
Setting it to an appropriate value can make search faster, if the difference
between the sides is known \textit{a priori}.
Our decision procedure also takes this an argument and passes it on to the CoRN tactic.

We have shown the intermediate step above for illustration purposes only.
The actual tactic proves the theorem in one step:

\begin{verbatim}
Example xkcd217 : (exp PI - PI < 20).
R_solve_ineq (1#1)%Qpos.
Qed.
\end{verbatim}

Automatic rewriting is not enough to convert partial functions like division and
logarithm. The additional parameters needed in CoRN are the domain conditions.
The tactic itself could be called recursively to generate the
assumptions.
Unfortunately Coq 8.1 cannot automatically rewrite inside dependent products, making the recursive tactic more difficult to create.
We understand that Coq 8.2's new setoid rewriting system will allow rewriting in dependent products, and we expect this to greatly simply the creation of a recursive tactic.

\subsection{Using facts about Coq reals in CoRN}

The standard library of Coq contains more properties of real numbers than CoRN. It
also contains more tactics, like \verb|fourier| for solving linear constraints.
By using the isomorphism the other way, it is possible to apply these tactics while working with CoRN.
Using the isomorphism this way is controversial because
using the classical reals means that the axioms of classical logic are assumed.

We will show how a goal that would normally be proved by the \verb|fourier| tactic
in Coq reals can be done in CoRN. We will show it on a very simple goal, but the
procedure is similar:

\begin{equation}
  x < y \Rightarrow x \le y + 1.
\end{equation}

The goal written formally in Coq is:

\begin{verbatim}
Goal forall x y:IR, (x [<] y) -> (x [<=] y [+] One).
\end{verbatim}

After introducing the assumptions we can apply the isomorphism to the inequalities
both in the assumptions and in the goal:

\begin{verbatim}
intros x y H; rapply IR_le_as_R_back.
assert (HH := R_lt_as_IR_back _ _ H).
\end{verbatim}

Since the isomorphism preserves all the functions in the goal and assumptions, we
can apply the facts to change the terms to terms operating on the isomorphism of the
variables, and then the \verb|fourier| tactic is applicable:

\begin{verbatim}
replace RHS with (IRasR y + IRasR One) by symmetry; rapply IR_plus_as_R.
replace (IRasR One) with 1. 2: symmetry; apply IR_One_as_R.
fourier.
Qed.
\end{verbatim}

A similar transformation can be performed to use other facts and tactics from the Coq library.


\section {Related Work}\label{p5:sec:related}
Melquiond has created a Coq tactic that can solve some linear inequalities over real number expressions using interval arithmetic and bisection~\cite{Mel08b}.
This tactic is currently limited to expressions from arithmetic operations and square root, but could support transcendental functions via polynomial approximations.  It also has the advantage that it can solve some problems that involve constrained variables.

Many other proof assistants include facts about transcendental functions
that could be used for approximating expressions that involve them. 
However there are few mechanisms for approximating real numbers 
automatically since to compute effectively this has to be done either by constructing the
real numbers with approximation in mind, or by using special features
of the proof assistant. The latter approach is used effectively for
example in \textsc{HOL Light}, due to a close interplay between syntax
and semantics.

The real numbers in \textsc{HOL Light} are constructed
\cite{harrison-thesis}, but the approximation mechanism does not
depend on this. Instead the \verb|calc_real| tactics use the fact that
terms are transparent and destruct a term or goal to look inside
it. This allows creating a tactic that approximates a given real
expression, by approximating subexpressions and is used to create the
\verb|REALCALC_REL_CONV| tactic that proves inequalities on ground
real numbers.

Obua developed a computing library for Isabelle \cite{isabelle}.
In his PhD \cite{obuaphd} he shows examples of computing bounds on
real number expressions using computation rather than deduction.

Lester implemented approximations of real number expressions in PVS~\cite{lester:2008}. Results of real number functions are proved to have fast converging Cauchy sequences when their parameters have fast converging Cauchy sequences.  Cauchy sequences for many real number functions are effective and can be evaluated inside PVS.

\section{Conclusion}\label{p5:sec:concl}

We have formalized a proof that the axioms of Coq's classical real numbers imply the decidability of $\Pi^1_0$ statements.
We used this fact to prove that these classical real numbers form a constructive real number structure.
Then we used the fact the all real number structures are isomorphic to use tactics designed for one domain to solve problems in the other domain.
In particular, we showed how to automatically prove a class of strict inequalities on real number expressions.

The lemmas showing the decidability of $\Pi^1_0$ statements have been added to the standard library and will be made available in the 8.2 release of Coq.
The isomorphism and the tactics used to prove inequalities over the Coq reals have been added to the CoRN library.  They will be available with the next release of CoRN, which will appear at the same time Coq 8.2 is released.

\subsection{Future Work}\label{p5:sec:future}

We wish to extend our tactics to solve inequalities over terms that involve partial functions.
This should be easier to do in Coq 8.2. Currently the translation of expressions from one domain to another is quite slow.
We would like to investigate ways that this could be made faster.
We would also like to automate the translation from CoRN expressions to Coq expressions so that CoRN can have its own \verb|fourier| tactic.


\bibliographystyle{plain}
\bibliography{P5}   

\begin{thebibliography}{10}

\bibitem{Coq:2007}
The {Coq} {Development}~{Team}.
\newblock {\em The {Coq} Reference Manual, version 8.1}, February 2007.
\newblock Distributed electronically at \url{http://coq.inria.fr/doc}.

\bibitem{cornsrc}
{Constructive Coq Repository at Nijmegen}, 2008.
\newblock \url{http://corn.cs.ru.nl/}.

\bibitem{luisthesis}
Lu\'is Cruz-Filipe.
\newblock {\em Constructive Real Analysis: a Type-Theoretical Formalization and
  Applications}.
\newblock PhD thesis, University of Nijmegen, April 2004.

\bibitem{lcf04-corn}
Lu\'{\i}s Cruz-Filipe, Herman Geuvers, and Freek Wiedijk.
\newblock {C-CoRN}, the constructive coq repository at nijmegen.
\newblock In Andrea Asperti, Grzegorz Bancerek, and Andrzej Trybulec, editors,
  {\em MKM}, volume 3119 of {\em Lecture Notes in Computer Science}, pages
  88--103. Springer, 2004.

\bibitem{cornhierarchy}
Herman Geuvers, Randy Pollack, Freek Wiedijk, and Jan Zwanenburg.
\newblock A constructive algebraic hierarchy in {C}oq.
\newblock {\em Journal of Symbolic Computation, Special Issue on the
  Integration of Automated Reasoning and Computer Algebra Systems},
  34(4):271--286, 2002.

\bibitem{harrison-thesis}
John Harrison.
\newblock {\em Theorem Proving with the Real Numbers}.
\newblock Springer-Verlag, 1998.

\bibitem{lester:2008}
David~R. Lester.
\newblock Real number calculations and theorem: Proving validation and use of
  an exact arithmetic.
\newblock In Otmane Ait-Mohamed, editor, {\em TPHOLs}, volume 5170 of {\em
  Lecture Notes in Computer Science}, pages 215--229. Springer, 2008.

\bibitem{Letouzey02}
Pierre Letouzey.
\newblock A new extraction for {C}oq.
\newblock In Herman Geuvers and Freek Wiedijk, editors, {\em TYPES}, volume
  2646 of {\em Lecture Notes in Computer Science}, pages 200--219. Springer,
  2002.

\bibitem{Mel08b}
Guillaume Melquiond.
\newblock Proving bounds on real-valued functions with computations.
\newblock In Alessandro Armando, Peter Baumgartner, and Gilles Dowek, editors,
  {\em Proceedings of the 4th International Joint Conference on Automated
  Reasoning}, Lectures Notes in Computer Science, Sydney, Australia, 2008.

\bibitem{isabelle}
Tobias Nipkow, Lawrence~C. Paulson, and Markus Wenzel.
\newblock {\em Isabelle/HOL - A Proof Assistant for Higher-Order Logic}, volume
  2283 of {\em Lecture Notes in Computer Science}.
\newblock Springer, 2002.

\bibitem{miladsthesis}
Milad Niqui.
\newblock {\em Formalising Exact Arithmetic: Representations, Algorithms and
  Proofs}.
\newblock PhD thesis, Radboud Universiteit Nijmegen, September 2004.

\bibitem{obuaphd}
Steven Obua.
\newblock {\em Flyspeck II: The Basic Linear Programs}.
\newblock PhD thesis, Technische Universitat Munchen, 2008.
\newblock submitted.

\end{thebibliography}

\end{document}